\documentclass{elsart}
\usepackage{latexsym}
\usepackage{amsmath}
\usepackage{amssymb}
\usepackage{amsfonts}

\def\cal{\mathcal}

\numberwithin{equation}{section}

\newcommand{\beq}{\begin{eqnarray}}
\newcommand{\eeq}{\end{eqnarray}}

\newcommand{\scbg}{semicentral bigroupoid}
\newcommand{\scbgs}{semicentral bigroupoids}

\newcommand{\Scbgs}{Semicentral bigroupoids}
\newcommand{\rs}{rectangular structure}
\newcommand{\rss}{rectangular structures}

\newcommand{\beginproof}{Proof:}
\newcommand{\bendproof}{\hfill\(\Box\)}

\begin{document}
\begin{frontmatter}

\title
{Towards a Noether--like conservation law theorem for 
one dimensional reversible cellular automata\thanksref{projthx}}
\author{Tim Boykett}
\address{Time's Up Research Department \\and\\
Department of Mathematics, Johannes--Kepler University, Linz, Austria}
\ead{tim@timesup.org}
\thanks[projthx]{Supported by BKA.Kunst, Land Ober\"osterreich and Linz.Kultur}
%\date{June 2003}
\renewcommand{\baselinestretch}{1.05}

\begin{abstract}
Evidence and results suggesting that a Noether--like theorem for
conservation laws in 1D RCA can be obtained. Unlike Noether's theorem,
the connection here is to the maximal congruences rather than the
automorphisms of the local dynamics.

We take the results of Takesue and Hattori (1992) on the
space of additive conservation laws in one dimensional cellular 
automata.
In reversible automata, we show that conservation laws correspond
to the null spaces of certain well-structured matrices.

It is shown that a class of conservation laws exist that correspond
to the maximal congruences of index 2. 
In all examples investigated, this is all
the conservation laws. 
Thus we conjecture that there is an equality
here, corresponding to a Noether--like theorem. 
\end{abstract}

\end{frontmatter}

%\maketitle

\section{Introduction}

This paper outlines some  investigations into the
conservation rules of one dimensional reversible cellular automata (RCA).
Using results from Takesue and Hattori, we obtained examples for
all RCA of low order ($\leq 9$). Based upon these observations,
we noted a strong connection with congruences, in particular maximal congruences.
We propose a conjecture relating the two, akin to Noether's result
about conservation laws in continuous systems.

Conservation laws in physical systems are of great help in interpreting
the properties of these systems. The investigation of 
conservation laws in models of
physical systems, i.e.\ cellular automata, seems thus relevant.
In particular, (microscopic) reversibility in physical systems is
often of  importance, leading us to investigate reversible
cellular automata.

\section{Reversible cellular automata}

One dimensional reversible cellular automata (1DRCA or simply RCA) are 
invertible, continuous mappings of $A^Z$ to itself 
that commute with the shift operator. $A$ is a finite set of cell states.
$Z$ can be either the integers or
the integers modulo $n$ for some $n$. The
metric measures the size of the matching middle, i.e.\ 
if $a_i = b_i$ for all $|i| < k$ and  $a_i \neq b_i$ for some $|i| = k$
then $d(a,b) = \frac{1}{2^k}$.
The shift operator $\sigma$ is defined by $\sigma(a)_i = a_{i+1}$.
CA can be defined by the set $A$ and the \emph{local rule} $f:A^n \rightarrow A$
\cite{richardson72}.
Our CA are binary (radius one half), the local rule
is a binary operation. If $f$ is the local
rule, then $F$ the global mapping, is defined by
$F(a)_i = f(a_i, a_{i+1})$.

 We need only consider a class of $(2,2)$--algebras
known as \emph{\scbgs} to investigate RCA \cite{pedersen92,boykett03}. 
These are defined as $(A,\bullet,\circ)$ with
the identities:
\beq
(a \circ b) \bullet (b \circ c ) &=& b \\
(a \bullet b) \circ (b \bullet a) &=& b
\eeq
It is relatively easy to show that all \scbgs\ are the composition of an
idempotent \scbg\ and a permutation, that is
$a \bullet b = \rho (a \bar\bullet b)$ where $\bar\bullet$ is an idempotent
\scbg\ operation on $A$ and $\rho$ is a permutation of $A$. It turns out
that $\rho$ is the square map of $\bullet$, i.e.\ $\rho a = a \bullet a$.

In general we can combine any \scbg\ $A$ with  a permutation $\rho$
in this way. We call the resulting \scbg\ the \emph{lifting} of
$A$ by $\rho$.

Examples of idempotent \scbgs\ include rectangular bands, which are the
only associative \scbgs. 

The following is an idempotent nonassociative \scbg\ of order six.
\begin{eqnarray*}
\begin{array}{ c |  c c c c c c }
\bullet & 1& 2& 3& 4& 5& 6\\
\hline
1 & 1 & 1 & 3 & 4 & 4 & 3\\
2 & 2 & 2 & 5 & 6 & 5 & 6\\
3 & 1 & 1 & 3 & 4 & 4 & 3\\
4 & 1 & 1 & 3 & 4 & 4 & 3\\
5 & 2 & 2 & 5 & 6 & 5 & 6\\
6 & 2 & 2 & 5 & 6 & 5 & 6\\
\end{array}
&\mbox{  }&
\begin{array}{ c |  c c c c c c }
\circ & 1& 2& 3& 4& 5& 6\\
\hline
1 & 1 & 2 & 1 & 1 & 2 & 2\\
2 & 1 & 2 & 1 & 1 & 2 & 2\\
3 & 3 & 6 & 3 & 3 & 6 & 6\\
4 & 4 & 5 & 4 & 4 & 5 & 5\\
5 & 3 & 5 & 3 & 3 & 5 & 5\\
6 & 4 & 6 & 4 & 4 & 6 & 6\\
\end{array}
\end{eqnarray*}

We will often omit the $\bullet$ symbol and use juxtaposition to
denote the operation where it causes no confusion.

The \scbg\ axioms are symmetric in the two operations. Thus
statements about $(A,\bullet)$ apply to $(A,\circ)$ equally. There is no
known equational definition of a \scbg\ using only one operation. There
is a combinatorial description of \scbgs\ which only requires
us to investigate one of the operations. As is the case in
lattices, if $(A,\bullet,\circ)$ and
$(A,\bullet,*)$ are \scbgs, then $(A,\circ)=(A,*)$ (not just isomorphic).

\Scbgs\ are \emph{rectangular},  $ab = cd \Rightarrow ad=cb=ab$.
Thus for any element $a \in A$ we obtain sets $L_a,R_a \subseteq A$ such that
$l\in L_a,\,r\in R_a \Leftrightarrow lr = a$. For any $a,b \in A$,
$|L_a| = |L_b|, |R_a| = |R_b|$ and $|L_a||R_a| = |A|$. This ordered pair
$(|L_a|,|R_a|)$ is called the \emph{format} of the \scbg $A$.

These pairs $\{(L_a,R_a) | a \in A\}$ form a combinatorial structure
known as a \emph{\rs}. 

\begin{defn}
A {\em Rectangular Structure} 
on a set $S$, called the {\em base set}, 
is a collection $\cal R$ of ordered pairs of subsets, 
called {\em rectangles},
of $S$, such that
\begin{eqnarray}
&\forall (s,t)\in S^2 \;\exists !\; R \in {\cal R} \mbox{ such that }
	(s,t) \in R \label{rs_axiom1}\\
&\forall R,Q \in {\cal R}, \vert R_1 \cap Q_2 \vert = 1 \label{rs_axiom2}
\end{eqnarray}
where we identify $R = (R_1,R_2) = R_1 \times R_2$.
\end{defn}

There is a one--to--one correspondence
between idempotent \scbgs\ and \rss.

There are some special classes of \rss. Two partitions $\Pi,\Theta$ of a set
are called \emph{orthogonal} if $\forall P \in \Pi, T\in \Theta$, 
$|P\cap T| = 1$.

\begin{defn}
A \emph{partitioned \rs} is defined by a set $S$, a partition
$\Pi$ of $S$ called the \emph{primary partition} and a collection
$\{\Theta_\pi: \pi \in \Pi\}$ of partitions of $S$ that are
orthogonal to $\Pi$.

The rectangular structure is then defined as $\{(\pi,T): \pi \in \Pi,\, T \in \Theta_\pi\}$
for a \emph{left--partitioned \rs} and $\{(T,\pi): \pi \in \Pi,\, T \in \Theta_\pi\}$
for a \emph{right--partitioned \rs}.
\end{defn}

It is simple to show that these satisfy the axioms (\ref{rs_axiom1}) and  (\ref{rs_axiom2}).
They have a simplified structure and special properties.
The example of order 6 above is left--partitioned with primary
partition $134|256$ and secondary partitions $12|36|45$ and $12|35|46$.

\section{Conservation Laws}

We are interested in conservation laws, which are numerical properties of
states of the RCA that do not change over time, i.e.\ with applications
of the global mapping.

An (additive) conservation law (\emph{conslaw}) is a mapping $\phi$ from
$A$ to the reals such that, if we define $\Phi(a) := \sum \phi(a_i)$ where this sum is defined,
then $\Phi(a)  = \Phi(Fa)$ where $F$ is the global mapping of the CA.

The set of conslaws for a given CA rule is a vector space over the reals. Thus the
problem is to find a basis for the vector space of conslaws for a given
rule.
The mapping $\phi_r:a\mapsto r$  taking all elements of $A$ to a given real $r$ is a trivial
conservation law. Thus we can force one element $o \in A$ to have
$\phi(o)=0$: if $\bar\phi$ is a conslaw, 
then $\phi(a) = \bar\phi(a) - \bar\phi(o)$
is a conslaw with this property.

Let us consider a CA with only two cells. The conslaw requirement then states
that 
\beq
\phi a + \phi b = \phi(ab) + \phi(ba)\, \forall a,b \in A \label{eq_twocell}
\eeq 
We call this the 
\emph{two--cell requirement}.

Hattori and Takesue \cite{hattoritakesue91} have demonstrated that the conslaws must
satisfy a simple equation for all $x,y \in A$:
\beq 
\phi(x) - \phi(ox) + \phi(oy) - \phi(xy) = 0  \label{eqn_ht}
\eeq

\begin{lem}
Let $\phi$ be a conslaw for a nonidempotent \scbg\ $(A,\cdot)$ with square
map $\rho$. Then 
$\phi$ will be constant on all orbits of $\rho$ and $\phi$ will be a
conslaw on the idempotent lifting of $A$.
Conversely, if $\phi$ is a conslaw of an idempotent \scbg\ $(A,*)$ and
$\phi$ is constant on the orbits of the permutation $\rho$, then
$\phi$ is a conslaw on the lifting of $A$ by $\rho$.
\end{lem}
\beginproof
For the first statement, take $a\in A$, then $\phi a = \phi \rho a$
by (\ref{eq_twocell}). So $\phi$ is constant on $\rho$--orbits. 

Let $\phi$ satisfy (\ref{eqn_ht}) on $(A,\cdot)$ and be
constant on orbits of the permutation $\rho$. 
Then $\phi(a*b)=\phi\rho^{-1}(a\cdot b) = \phi(a\cdot b)$
so $\phi$ satisfies (\ref{eqn_ht}) on $(A,*)$ and is a conslaw.

Thus the rest of the forward argument and the converse statement
can be seen to be true. 
\bendproof

Thus we need only concern ourselves with idempotent
\scbgs\ for the rest of this paper.

The equations (\ref{eqn_ht}) can be formulated as a linear algebra problem.
We label the elements of $A$ as $\{1,\ldots,n\}$, then a
matrix $M$ is defined so that if $Mv = 0$ then $\phi(i) = v_i$
is a conslaw. The columns are indexed by the elements of $A$, the
rows correspond to the pairs $(x,y) \in A^2$ that give us the equation.

\begin{lem}
The entries of the matrix $M$ are $-1,0,1$. 
The row and column sums are equal to $0$.
\end{lem}
\beginproof
For an entry of $M$ to be $2$, we require $x = oy$. Thus $xx=oy$ so
by the rectangular property, $xx=xy=ox=oy$ and  (\ref{eqn_ht}) becomes $0=0$.
Similarly if an entry of $M$ is to be $-2$, we require $ox=xy$. By
the rectangular property, $ox=oy=xx=xy$ so (\ref{eqn_ht}) reduces to $0=0$.

The rows of the matrix thus consist of either all $0$s, or exactly
one $1$ and one $-1$ or exactly two $1$s and two $-1$s. 
The row sum is always equal to $0$.

In order to calculate the column sum, we count the occurrences
of a given element $z$ in pairs $x,y$. We  see that $z$ occurs
equally often as $ox$ and $oy$ by symmetry, and equally often ($|A|$ times) 
as $x$ (the pairs $(z,y)$) and as $xy$ (the 
rectangle of pairs  $(x,y)$ such that $xy=z$). Thus the sum is zero.
\hfill$\Box$

With these results it has been possible to calculate the space of conslaws
for all 1DRCA with $A$ up to order 9, using the exhaustive
generation results explained in \cite{boykett03}. Tests have been performed
with randomly generated examples of orders 12 and 16. We will return to the
results of these tests later.

\section{Morphisms and congruences}

If $\psi: A \rightarrow B$ is a \scbg\ morphism, and $\phi$
is a conslaw for $B$, then $\phi\circ\psi$ is a conslaw for $A$.
We call these \emph{pullbacks} of conservation laws.

Define $K(A)$ as follows: if $A$ is simple, then $K(A)$ is the
set of conservation laws of $A$, otherwise, it is the set of
pullbacks of $K(\bar A)$ for each homomorphic image $\bar A$  of $A$.
We obtain the following.

\begin{lem}
$K(A)$ is defined by the maximal congruences, i.e.\ the simple
images of $A$ only.
\end{lem}

Thus, given an arbitrary RCA, we can obtain a class of conservation
laws by looking only at the factors modulo the maximal congruences,
i.e.\ the simple images of the algebra.

The question arises as to simple \scbgs. The known examples
are the left and right constant \scbgs\ of order 2 and two examples
of order 9. There is no reason to believe that there are not more.
However, the following results show that they do not have a 
particularly simple structure.

\begin{defn}
A \rs\ $\cal R$ on a set $A$ is \emph{left (right) partitionable} if there exists
a nontrivial partition $\Pi$ of $A$ such that $\forall R \in {\cal R},\, \forall a\in R_1:
 R_1 \subseteq [a]_\Pi$ ($\forall R \in {\cal R},\, \forall a\in R_2:
 R_2 \subseteq [a]_\Pi$).
\end{defn}

\begin{lem}
Let $\cal R$ be a \rs\ on the set $A$. If  $\cal R$ is left (right) 
partitionable then ${\cal R}/\Pi$ is
a \rs\ on $A/\Pi$ with format $1 \times |\Pi|$ $(|\Pi| \times 1)$.
\end{lem}
\beginproof
We concentrate on the left partitionable case. 
Fix $R \in {\cal R}$. It is clear that $|R_1 / \Pi| = 1$.
For all $ a\in A$ s.t.\ $[a] \in R_2/\Pi,\, \exists Q \in {\cal R}$ such that $a\in Q_1$.
Then $R_2 \cap Q_1 \neq \emptyset \Rightarrow [a]\in R_2/\Pi$, i.e.\ $R_2/\Pi = A/\Pi$.
Thus we have a $1 \times |\Pi|$ \rs. A similar argument demonstrates the
result for right partitionable \rss.
\bendproof

Define a \rs\ to be \emph{simple} if the corresponding \scbg\ is simple.

\begin{cor}
Partitioned \scbgs\ are simple iff of order 2.
\end{cor}
This follows as partitioned implies partitionable.
\begin{cor}
A \scbg\ of format $2 \times n$ or $n \times 2$ is simple iff $n=1$.
\end{cor}
\beginproof
Without loss of generality we consider format $2 \times n$.
Suppose $n \neq 1$.
The left side of a $2\times n$ rectangle is a pair. We have $2n$ pairs
on $2n$ points in which every point appears in exactly two pairs.
Thus we have a  graph which is a union of cycles. This is partitionable
(and thus not simple)
unless there is a unique cycle covering all $2n$ points.

We label the elements of $A = \{1,2,\ldots,2n\}$, in order along this
cycle. Take some rectangle $R$. If the unique $a \in R_1 \cap R_2$
is even, then $R_2$ consists of only even elements, as it must
contain $n$ elements and no two may be adjacent on the cycle by the 
unique intersection property. Similarly
if $a$ is odd, then $R_2$ consists only of odd elements. 
Thus the \rs\ is right partitioned, and not simple.
\bendproof

Thus a simple idempotent \scbg\ of order higher than 2 must have order at
least $3^2$, and it turns out that there are precisely two such simple
\scbgs\ by investigating the exhaustive lists generated in \cite{boykett03}.

The simple \scbgs\ of order 2 have the conslaw that 
maps one element to 0, the other
to 1. The simple examples of order 9 have no nontrivial conslaws.

\begin{lem}
A rectangular structure is partitionable iff the idempotent \scbg\ can
be mapped to a \scbg\ of order 2.
\end{lem}
\beginproof
The forward direction is clear, as the partition forms a congruence
in the \scbg. We can collapse any partition into a two--class partition to
obtain a two element image.

For the reverse direction we use the partition generated by the
congruence classes of the homomorphism. Since the image has
two elements, wlog the corresponding \rs\ is of format $1\times 2$.
Thus $R_1$ lies within one class of the partition for all rectangles $R$.
Thus the \rs\ is partitionable.
\bendproof

\section{0,1 conslaws}

All calculated examples to date have a conslaw basis with $\{0,1\}$--vectors,
that is, the vectors contain only entries from the set $\{0,1\}$.
The following results show that this means that all known conslaws are
$K(A)$ type.

\begin{lem}
\label{lem_ac}
Let $\phi:A\rightarrow \{0,1\}$ be a conslaw. If $\phi a = 0,\,\phi b = 1$ and
$\phi(ab)=0$ then $\phi(ac)=0$ and $\phi(bc)=1$ for all $c\in A$.
\end{lem}
\beginproof
%Assume there is some $c$ such that $\phi(ac)=1$.
%By the two--cell requirement, $\phi a + \phi c =\phi c = \phi(ac)+\phi(ca) \geq 1$
%so $\phi c = 1$.
By the Hattori--Takesue equation with $o=a,x=b,y=c$ we have
\beq
0 &=& \phi b - \phi(ab)+\phi(ac) - \phi(bc) \\
 \phi(bc)-\phi(ac) &=& 1 \\
 \phi(bc) = 1 &\mbox{and }& \phi(ac)= 0
\eeq
\bendproof
\begin{lem}
\label{lem_ca}
Let $\phi:A\rightarrow \{0,1\}$ be a conslaw. If $\phi a = 0,\,\phi b = 1$ and
$\phi(ab)=1$ then $\phi(ca)=0$ and $\phi(cb)=1$ for all $c\in A$.
\end{lem}
\beginproof
First note that by the hypothesis, $\phi(ba)=0$ using the two--cell requirement (\ref{eq_twocell}).
Assume $\phi(ca) = 1$. Then  by the two--cell requirement,
\[ \phi a + \phi c = \phi c = \phi (ca) + \phi(ac) \geq 1 \Rightarrow \phi c = 1
\]
Furthermore 
\[ 2=\phi b + \phi c = \phi (bc) + \phi(cb) \Rightarrow \phi(bc)=\phi(cb) = 1
\]
Then by Hattori and Takesue we obtain
\[0 = \phi c - \phi (bc) + \phi (ba) - \phi(ca) = 1 - 1+0-1\]
which is a contradiction,
thus $\phi(ca)=0$. From the two--cell requirement 
\[\phi a + \phi c = \phi c = \phi(ac) + \phi(ca) = \phi(ac)\]
so the Hattori--Takesue equation
\[\phi c - \phi(ac) + \phi(ab) -\phi(cb) = \phi(ab)-\phi(cb) = 0\]
implies that $\phi(cb)=\phi(ab)=1$ and we are done.
\bendproof

\begin{thm}
If $\phi:A \rightarrow \{0,1\}$ is a conslaw, then $\phi$ is a
morphism of $A$ onto a \scbg\ of order 2.
\end{thm}

\beginproof
Fix some $a,b \in A$ such that $\phi a = 0$ and $\phi b = 1$.
Define an operation $*$ on $\{0,1\}$ by
\begin{eqnarray*}
\begin{array}{ c |  c c }
* & 0& 1\\
\hline
0 & 0 & \phi(ab) \\
1 & \phi(ba) & 1\\
\end{array}
\end{eqnarray*}   

We claim that $\phi:(A,\bullet) \rightarrow (\{0,1\},*)$ is a morphism.
By the two--cell requirement, exactly one of $\phi(ab)$ and $\phi(ba)$ is
$0$ and the other is $1$.

Case 1: $\phi(ab)=0$. In this case $x*y=x$.
By Lemma \ref{lem_ac} we know that $\phi(ac)=0$ and $\phi(bc)=1$ for all $c\in A$. 
If $\phi c = 1$ then replacing $a,b,c$ with
$a,c,d$ in the same Lemma, we see that $\phi(cd)=1$ for all $d \in A$,
so $\phi(cd) = \phi(c)*\phi(d)$ and $\phi$ is a morphism.
If $\phi c = 0$ then by the two--cell requirement 
$\phi b + \phi c = 1 = \phi(bc) + \phi(cb) \Rightarrow \phi(cb)=0$.
Replacing $a,b,c$ with $c,b,d$ in Lemma \ref{lem_ac} we obtain $\phi (cd)=0$
so $\phi(cd) = \phi c * \phi d$ and $\phi$ is a morphism.

Case 2: $\phi(ab)=1$. In this case $x*y=y$.
By Lemma  \ref{lem_ca} we know that $\phi (ca)=0$ and $\phi(cb)=1$  for all $c\in A$.
If $\phi c = 1$, then $\phi(ac) = \phi a + \phi c - \phi(ca) =1$ so 
replacing $a,b,c$ with $a,c,d$ in Lemma  \ref{lem_ca}
implies $\phi(dc) =1= \phi d * \phi c$, so $\phi$ is a morphism.
If $\phi c = 0$ then replacing $a,b,c$ with $c,b,d$ in Lemma  \ref{lem_ca}
gives us $\phi(dc)=0=\phi d * \phi c$  and $\phi$ is a morphism.
\bendproof

Thus if we can find a conslaw space with no $\{0,1\}$--basis, then we have
something special. Otherwise we find:

\begin{conj}
There is a one-to-one correspondence between the basis of
the space of nontrivial conservation laws of a CA rule and the maximal 
congruences of $A$ with factor of size 2.
\end{conj}

This result is similar to the
result from classical continuous dynamical systems that connects the
space of automorphisms of the system with its conserved quantities.
Note that examples show that a similar
result doesn't apply for the group of automorphisms
of the algebra. This was my starting point and it doesn't 
work except in the trivial cases.

Note that if this conjecture is true, we obtain that the following
concepts are equivalent:
\begin{itemize}
\item Additive conservation laws in one dimensional reversible cellular automata
\item Maximal congruences on \scbgs\ with two element factors
\item Partitionability of \rss.
\end{itemize}
This three--way connection seems a little strange.
It would imply that the dimension of 
the conslaw vector space is constant,
independent of which field we take for it. 

\section{Conclusion}

We investigated the properties of additive conservation laws in
one dimensional reversible cellular automata with a binary local rule.
We have shown that a class of such conservation laws exist,
determined by the maximal congruences of index two in the 
groupoid defined by the local rule.
Exhaustive testing of examples has shown that only such
laws exist. We demonstrated that these conservation laws
correspond exactly to those with a 0,1 basis.

We conjecture that this holds in general. Two results would
be of interest. Showing that no nontrivial conservation
laws exist on simple \scbgs\ of order greater than two would
strengthen but not prove the conjecture.
More important would be to determine that all nullspaces of
the conservation law defining matrix have a 0,1 basis.
In fact, limited experiments in general binary local rule
cellular automata indicate that this might be a general result:
all conservation laws lie in a space with a basis determined
by the maximal congruences of order two in the groupoid.

\section{Acknowledgments}

The work here started during a visit to Cris Moore at the Sante Fe institute.
After working with him for some time, we realised that Hattori and Takesue
had obtained all our results years before!

\bibliographystyle{alpha}
\bibliography{tims}

\begin{thebibliography}{Boy03}

\bibitem[Boy03]{boykett03}
Tim Boykett.
\newblock Efficient exhaustive listings of reversible one dimensional cellular
  automata.
\newblock Accepted, in press, 2003.

\bibitem[HT91]{hattoritakesue91}
Tetsuya Hattori and Shinji Takesue.
\newblock Additive conserved quantities in discrete-time lattice dynamical
  systems.
\newblock {\em Physica D}, 49:295--322, 1991.

\bibitem[Ped92]{pedersen92}
John Pedersen.
\newblock Cellular automata as algebraic systems.
\newblock {\em Complex Systems}, 6:237--250, 1992.

\bibitem[Ric72]{richardson72}
D.~Richardson.
\newblock Tesselations with local transformations.
\newblock {\em J. Comp. Sys. Sci.}, 6:373--388, 1972.

\end{thebibliography}

\end{document}